\documentclass[%
 reprint,
 aps,
 showpacs,
 showkeys,
 nofootinbib,
 amsmath, 
 amssymb, 
 amsth, 
 amsfonts, 
 mathtools
]{revtex4-1}

\usepackage[final]{pdfpages}
\usepackage{hyperref}
\usepackage{graphicx}
\usepackage{dcolumn}
\usepackage{bm}

\begin{document}

  \title{New High-Pressure Phases of MoSe$_2$ and MoTe$_2$}

  \author{Oto Kohul\'{a}k}%
   \email[Corresponding author: ]{kohulak@fmph.uniba.sk}%
   \affiliation{Department of Experimental Physics, Faculty of Mathematics, Physics and Informatics, Comenius University in Bratislava, Mlynsk\'{a} dolina F2, 84248 Bratislava, Slovakia}%
  \author{Roman Marto\v{n}\'{a}k}%
   \affiliation{Department of Experimental Physics, Faculty of Mathematics, Physics and Informatics, Comenius University in Bratislava, Mlynsk\'{a} dolina F2, 84248 Bratislava, Slovakia}%

\date{\today}

  \begin{abstract}    
    
Three Mo-based transition metal dichalcogenides MoS$_2$, MoSe$_2$ and MoTe$_2$
share at ambient conditions the same structure 2H$_c$, consisting of layers where Mo atoms are surrounded by six chalcogen atoms in trigonal prism coordination. The knowledge of their high-pressure behaviour is, however, limited, particularly in case of MoSe$_2$ and MoTe$_2$. The latter materials do not undergo a layer-sliding transition 2H$_c$ $\rightarrow$ 2H$_a$ known in MoS$_2$ and currently no other stable phase besides 2H$_c$ is known in these systems at room temperature. Employing evolutionary crystal structure prediction in combination with \textit{ab initio} calculations we study the zero-temperature phase diagram of both materials up to Mbar pressures. We find a tetragonal phase with space group \textit{P4/mmm}, previously predicted in MoS$_2$, to become stable in MoSe$_2$ at 118 GPa. In MoTe$_2$ we predict at 50 GPa a transition to a new layered tetragonal structure with space group \textit{I4/mmm}, similar to CaC$_2$, where Mo atoms are surrounded by eight Te atoms. The phase is metallic already at the transition pressure and becomes a good metal beyond 1 Mbar. We discuss chemical trends in the family of Mo-based transition metal dichalcogenides and suggest that MoTe$_2$ likely offers the easiest route towards the post-2H phases.

  \end{abstract}
  

  \maketitle

\section{Introduction}
  
Transition metal dichalcogenides (TMD) are mostly layered materials with composition MX$_2$, where M is transition metal like W, Mo, Ti, Nb, Ta, etc. and X is a chalcogen such as S, Se, Te. While studied since long ago\cite{wilson1969,gmelin1995} they are currently of large interest, recently partly motivated also by the possibility of preparing graphene-like monolayers Refs.\cite{mak2010,chang2011,radisavljevic2011,wang2012}. Their bulk phases exhibit a broad spectrum of electronic behaviour, from insulator to semiconductor to superconductor \cite{wang2012}. Besides that they exhibit also rich structural diversity, which is related to two possibilities for local sixfold coordination (trigonal prism or octahedral), number of layers in the unit cell and stacking order, resulting in number of possible structures, such as 1T, 2H, 3R, etc\cite{wilson1969}. It is well known that their electronic properties can be modified by doping\cite{somoano1973}. Another important instrument allowing to tune their properties is pressure which can induce structural as well as electronic transitions.
  
In this study we focus on Mo-based TMDs which include MoS$_2$, MoSe$_2$ and MoTe$_2$. At ambient conditions all three compounds adopt a stable hexagonal 2H$_c$ structure, where Mo atoms are in trigonal prism coordination\cite{wilson1969}. This form is semiconducting with an indirect gap around 1 eV. Besides this form, MoS$_2$ and MoSe$_2$ are also known in metastable 3R form which differs from 2H$_c$ only by layer stacking. On the other hand, MoTe$_2$ can be prepared also in metallic metastable $\beta$-MoTe$_2$ monoclinic structure (space group \textit{P2$_1$/m}) with distorted octahedral layers\cite{brown1966}. Unlike 2H$_c$ (also called $\alpha$-MoTe$_2$) this form is semimetallic at ambient pressure. Besides the $\beta$-form also an orthorombic T$_d$ structure with distorted layers is known\cite{clarke1978}.
  
Among the three compounds, the most studied one under pressure is MoS$_2$. 
It has been studied experimentally up to 38.8 GPa in Ref.\cite{aksoy2006} where an unexplained structural transition was found around 20 GPa. The results of this study were theoretically explained in Ref.\cite{hromadova2013} by proposing a phase transition consisting of layer sliding. This interpretation was experimentally confirmed in Refs.\cite{bandaru2014,chi2014} where the system was studied up to 51 GPa and 81 GPa, respectively. In the latter study MoS$_2$ was also found to undergo a metallization transition resulting from band overlap between 30 - 40 GPa, in agreement with theoretical prediction in Ref.\cite{hromadova2013}. The behaviour at even higher pressure in the Mbar range was recently studied experimentally\cite{chi2015} and theoretically in Ref.\cite{kohulak2015}. While the theoretical work\cite{kohulak2015} predicted a phase transition to a new tetragonal phase with space group \textit{P4/mmm} at 138 GPa or a chemical decomposition at 135 GPa to MoS and elemental sulphur, experimentally no structural transition was observed up to 200 GPa. The discrepancy can most likely be attributed to the slow kinetics at high pressure conditions.

In Ref.\cite{riflikova2014} the possibility of a layer sliding transition, similar to that in MoS$_2$, was studied theoretically in MoSe$_2$ and MoTe$_2$. It was shown that in these materials the 2H$_a$ structure is actually less stable then the 2H$_c$ one, due to larger volume, and therefore no sliding transition takes place. The systems were predicted to metallize at 40 and 19 GPa, respectively. This result for MoSe$_2$ was confirmed in Ref.\cite{zhao2015} together with the lack of sliding transition up to 60 GPa. In Ref.\cite{riflikova2014}, however, no structural search was performed in order to identify possible transitions to other structures in MoSe$_2$ and MoTe$_2$. Nothing is therefore known about the behaviour of MoSe$_2$ beyond 60 GPa, neither theoretically nor experimentally. In case of MoTe$_2$ a lot of experimental activity was recently devoted to study of its properties at ambient pressure. Main interest was given to its orthorhombic phase T$_d$ which was proposed\cite{soluyanov2015} and more recently proven to be a type II Weyl semimetal\cite{Qi2016}. Possible transitions to other stable structures at high pressure, however, remain unknown.

In this paper we aim at filling the gap in understanding of the high-pressure behaviour of MoSe$_2$ and MoTe$_2$ up to Mbar pressures. By applying evolutionary algorithms in combination with \textit{ab initio} calculations we predict new phases and analyze their properties. The paper is organized as follows. In section \ref{methods} we present the \textit{ab initio} and genetic algorithms approach we used. In section \ref{results} we present our predictions for both materials and analyze their properties. In the final section \ref{conclusions} we discuss the results and draw conclusions.   
  
\section{Methods}
\label{methods}

For our evolutionary search we have used the XtalOpt package\cite{lonie2011} in combination with \textit{ab initio} structural optimization implemented in \textsc{Quantum ESPRESSO}\cite{giannozzi2009} and VASP\cite{kresse1993} packages. We employed the projector augmented wave (PAW) method\cite{blochl1994,kresse1999} and the exchange-correlation energy was calculated with the PBE functional\cite{perdew1996}. Similarly to Ref. \cite{hromadova2013}, also here we decided not to apply vdW corrections in structural relaxations at high pressure (including the evolutionary searches). We compare in Fig. 2 in Supplementary Material our theoretical predictions for the lattice parameter $c$ in MoSe$_2$ with the experimental data from Ref. \cite{aksoy2008}. It can be seen that at pressures beyond 15 GPa the PBE calculation agrees very well with the experimental data. For calculation with VASP we have used PAW pseudopotentials (PP) with 14, 6, and 6 valence electrons for Mo, Se and Te, respectively. For calculation with \textsc{Quantum ESPRESSO} we have used PAW PP with the same number of valence electrons for Mo and Se but we used 16 valence electrons for Te.

  For enthalpy calculations we have used the VASP package\cite{kresse1993} where the energy cutoff was set to 400 eV. K-point grid for Monkhorst-Pack scheme\cite{monkhorst1976} was generated with ASE package\cite{bahn2002} with desired density of 7 points$/\AA^{-1}$. Electronic band structure calculations were performed with {\sc Quantum ESPRESSO}\cite{giannozzi2009} package and the spin-orbit coupling was neglected. We have used cutoff 80 Ry and K-point grid $18 \times 18 \times 6$ for selfconsistent calculation and $30 \times 30 \times 10$ for non-selfconsistent calculation for electron density of states. For phonon calculations we have used the {\sc Quantum ESPRESSO} package\cite{giannozzi2009} employing the density functional perturbation theory (DFPT). The energy cutoff for plane waves was chosen 70 Ry, grid of q-points was chosen $9\times9\times3$ in both materials.
  
We have performed various evolutionary searches with up to four formula units of MoSe$_2$ up to 150 GPa and up to eight formula units of MoTe$_2$ up to 110 GPa (for more details about search settings see supplemental material\cite{Sup_mat}). For search in MoSe$_2$ we have used as a starting population only randomly generated structures. In case of MoTe$_2$ at 50 GPa we have also included some known structures such as hexagonal $\alpha$-MoTe$_2$ - 2H$_c$, $\beta$-MoTe$_2$ - 1T$^\prime$ and orthorhombic T$_d$ while at 110 GPa we included 2H$_c$ and the already known \textit{I4/mmm} in order to speed up the calculations. The number of structures generated in all searches performed are shown in Table 1 in Supplemental material\cite{Sup_mat}. Our results are valid within the usual limitations of structural search, in particular the limited number of atoms in the simulation cell and the limited number of generated structures. 

  \begin{figure}[ht]
    \centering
    \includegraphics[width=0.45\textwidth]{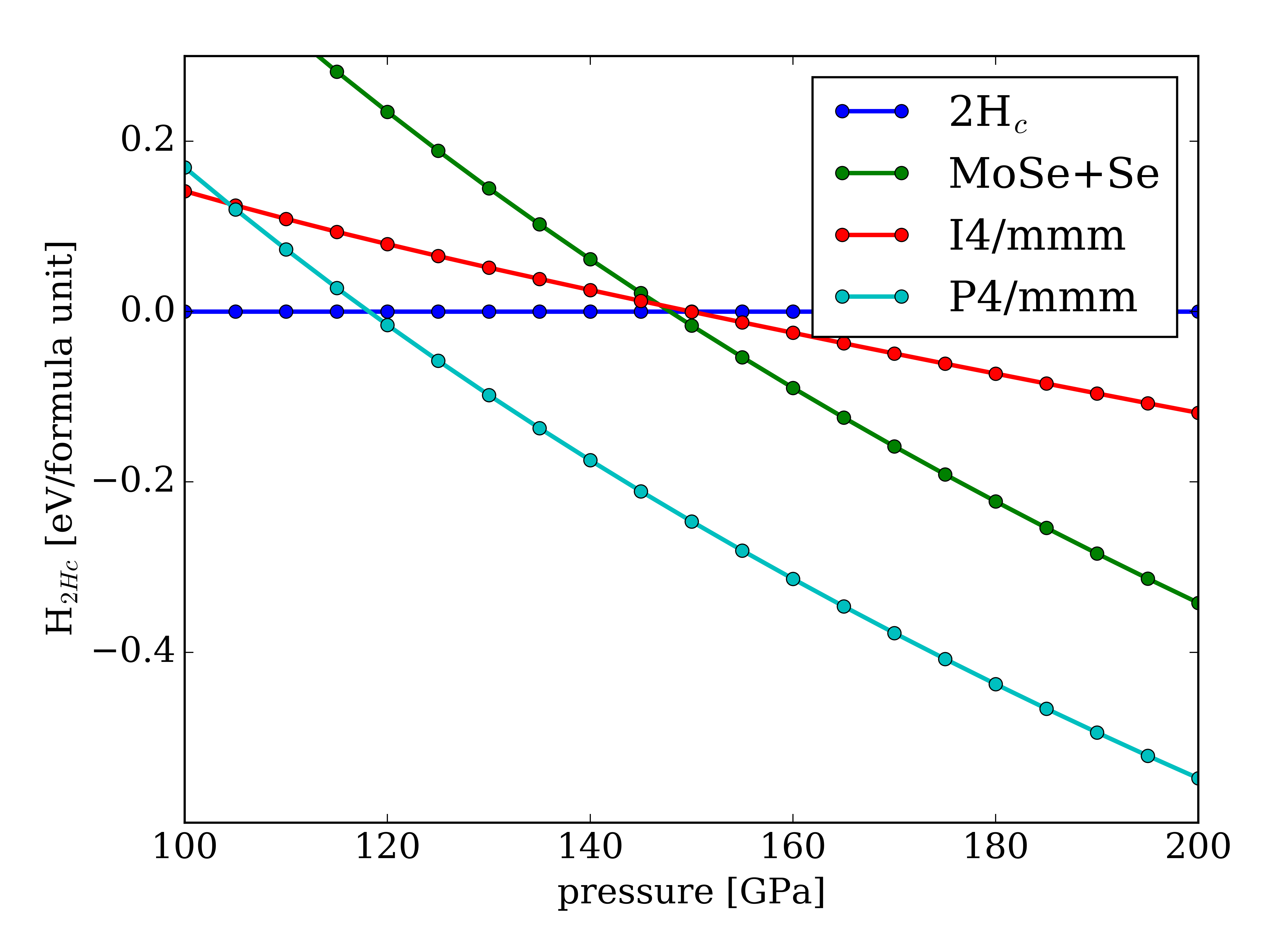}
    \caption{Enthalpy as function of pressure for MoSe$_2$ structures (calculated with VASP). Enthalpies are relative to the 2H$_c$ phase.}
    \label{fig:entcomp_decomp_MoSe2}
  \end{figure}

\section{Results}

\label{results}

\subsection{MoSe$_2$}

\begin{figure}[ht]
    \centering
    \includegraphics[width=0.45\textwidth]{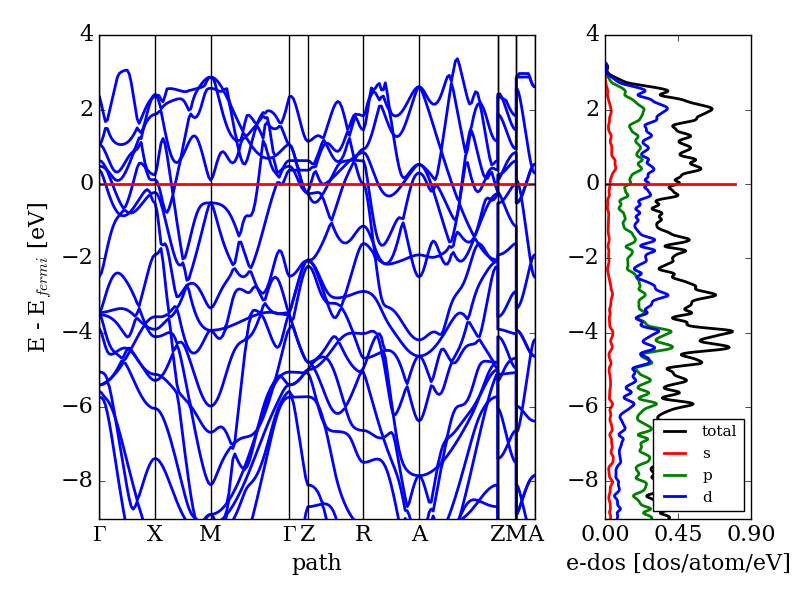}
    \caption{Band structure and projected density od states of MoSe$_2$ - \textit{P4/mmm} phase at 118 GPa. Last two sections are paths between X $\rightarrow$ R and M $\rightarrow$ A.}
    \label{fig:MoSe2_P4mmm_bandstructure}
  \end{figure}

In our structural search we found several known as well as new structures. The enthalpies of the relevant low-enthalpy ones as function of pressure are shown in Fig.\ref{fig:entcomp_decomp_MoSe2}. The ambient pressure layered phase 
2H$_c$ remains stable up to remarkably high pressure of 118 GPa. At 118 GPa a tetragonal \textit{P4/mmm} (No.123) phase, analogous to that predicted in Ref.\cite{kohulak2015} for MoS$_2$ becomes more stable, accompanied by decrease of volume by 3.7 \%.  At 120 GPa the lattice parameters are $a = 2.854$ \AA~and $c = 8.590$ \AA, Mo atoms are on Wyckoff positions 2h (0.5, 0.5, 0.166) and Se atoms are on 1d (0.5, 0.5, 0.5), 2g (0.0, 0.0, 0.321) and 1a (0.0, 0.0, 0.0) positions. This new phase is a good metal as can be seen in Fig.\ref{fig:MoSe2_P4mmm_bandstructure}, where the electronic band structure and the density of states are plotted.

\begin{figure}[ht]
    \centering
    \includegraphics[width=0.45\textwidth]{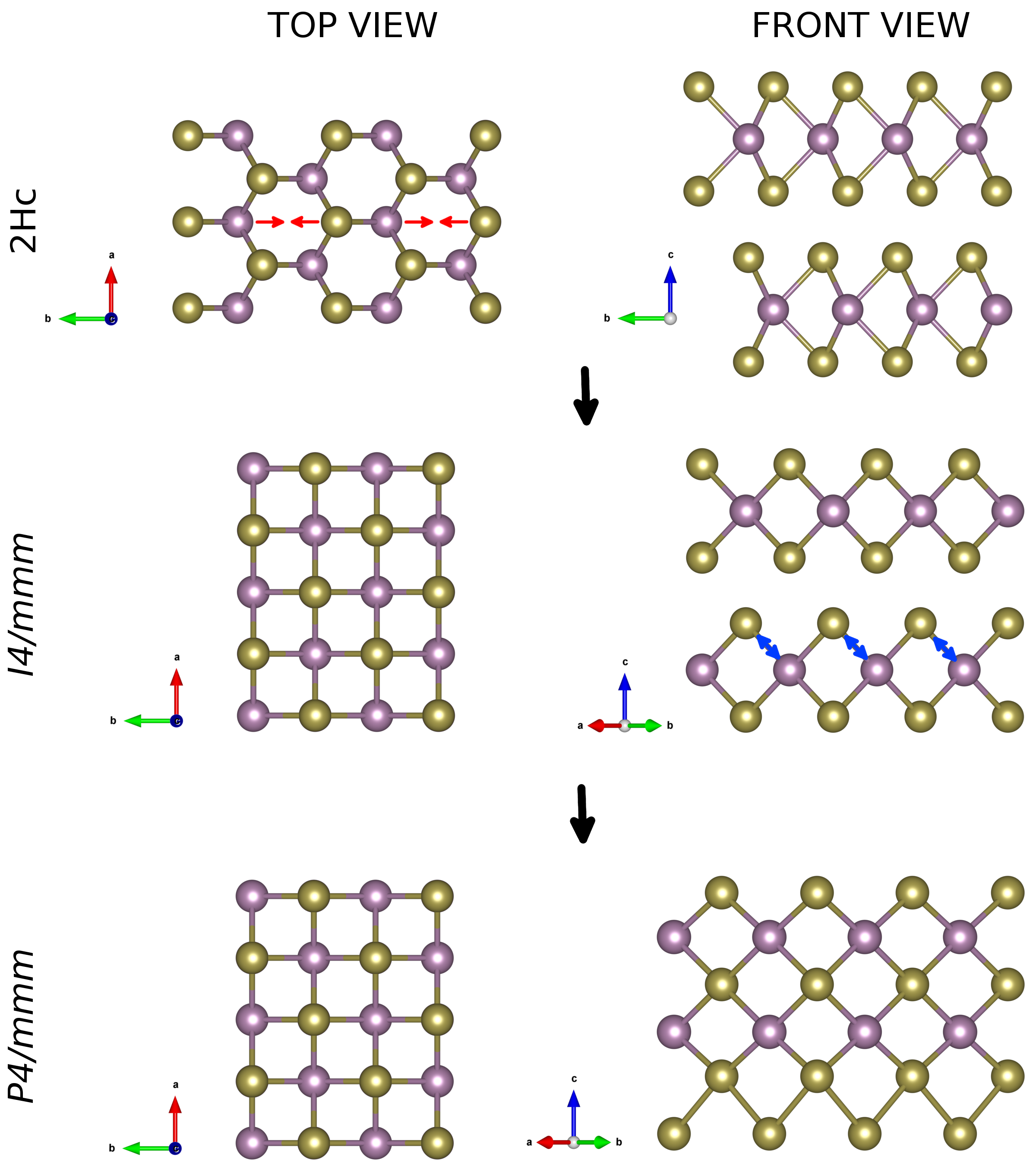}
    \caption{Tentative phase transition mechanism between 2H$_c$ (top panel) and \textit{I4/mmm} structures (middle panel) could proceed via uniaxial compression as schematically indicated by red arrows. The other speculated phase transition between \textit{I4/mmm} and \textit{P4/mmm} structures (bottom panel) can occur by exchange of positions between Mo and chalcogen atoms (indicated by blue arrows). Structures were visualised by \textsc{VESTA} package\cite{momma2011}.}
    \label{fig_I4mmm}
  \end{figure}

Assuming that direct transition to the \textit{P4/mmm} phase might be difficult due to kinetic reasons, as discussed in Ref.\cite{kohulak2015}, it is justified to consider possible transitions to metastable phases that compete with 2H$_c$ at pressures beyond 118 GPa. At 150 GPa there are two phases that nearly simultaneously become more stable than 2H$_c$. One of them represents chemical decomposition to MoSe + Se, where MoSe adopts the CsCl structure, again similarly to the case of MoS$_2$ \cite{kohulak2015} and Se the Se-VI structure\cite{degtyareva2005}. Since the structure of MoSe at conditions of the phase transition was unknown, an additional evolutionary search was performed and indeed confirmed the CsCl structure. For similar kinetic reasons as in case of \textit{P4/mmm} we believe that such decomposition might not be easily observed, at least at room temperature. A plausible scenario therefore appears to be a transition to a tetragonal phase with space group \textit{I4/mmm} (No.139) which becomes stable with respect to 2H$_c$ at 150 GPa. This phase also represents a layered structure (Fig.\ref{fig_I4mmm}), however, Mo atom now has an eightfold rather than sixfold coordination. While the structure is somewhat similar to CaC$_2$ (the same space group), the
lattice parameters ratio $\frac{c}{a}$ is substantially higher in \textit{I4/mmm}-MoSe$_2$ and therefore in contrast to CaC$_2$ the new phase has a layered character. To our knowledge such structure has never been observed in transition metal dichalcogenides. The phase transition is accompanied by a small volume drop of 1.0 \%. Structural parameters of the \textit{I4/mmm} phase at 150 GPa (equilibrium condition with 2H$_c$ phase) are $a = 2.685$ \AA~and $c = 9.454$ \AA~with Mo atom at Wyckoff position 2a (0., 0., 0.) and Se atom at 4e (0.5,  0.5, 0.845). It appears plausible to assume that this structure could be created from 2H$_c$ via a simple uniaxial compression or shear bringing two more chalcogen atoms in the vicinity of Mo atom as shown schematically in Fig.\ref{fig_I4mmm} (see also the next subsection \ref{MoTe2}). Since such mechanism does not require diffusion of atoms, the \textit{I4/mmm} phase could possibly be more easily kinetically accessible compared to both \textit{P4/mmm} phase and chemical decomposition. We thus predict it to be a likely candidate for the experimental outcome of compression of 2H$_c$ beyond 150 GPa even though it is only metastable with respect to both \textit{P4/mmm} and decomposition. We note that this structure was also found for MoS$_2$ in the study in Ref.\cite{kohulak2015}, however, in that case it has enthalpy higher than 2H$_a$ at 150 GPa by 0.51 eV per formula unit and therefore we did not consider it further.
  
\begin{figure}[ht]
    \centering
    \includegraphics[width=0.45\textwidth]{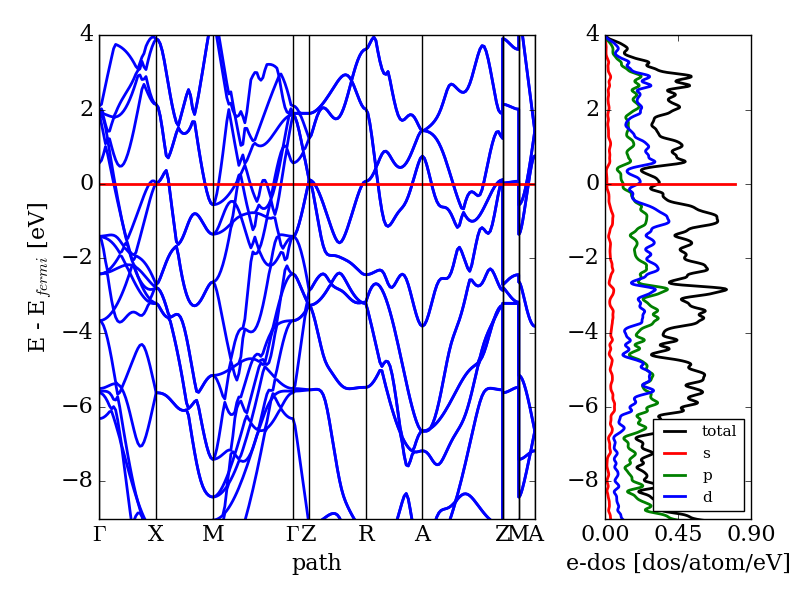}
    \caption{Band structure and projected density of states of the MoSe$_2$ - \textit{I4/mmm} phase at 150 GPa. Last two sections are paths between X $\rightarrow$ R and M $\rightarrow$ A.}
    \label{fig:MoSe2_I4mmm_bandstructure}
  \end{figure}

\begin{figure}[ht]
    \centering
    \includegraphics[width=0.45\textwidth]{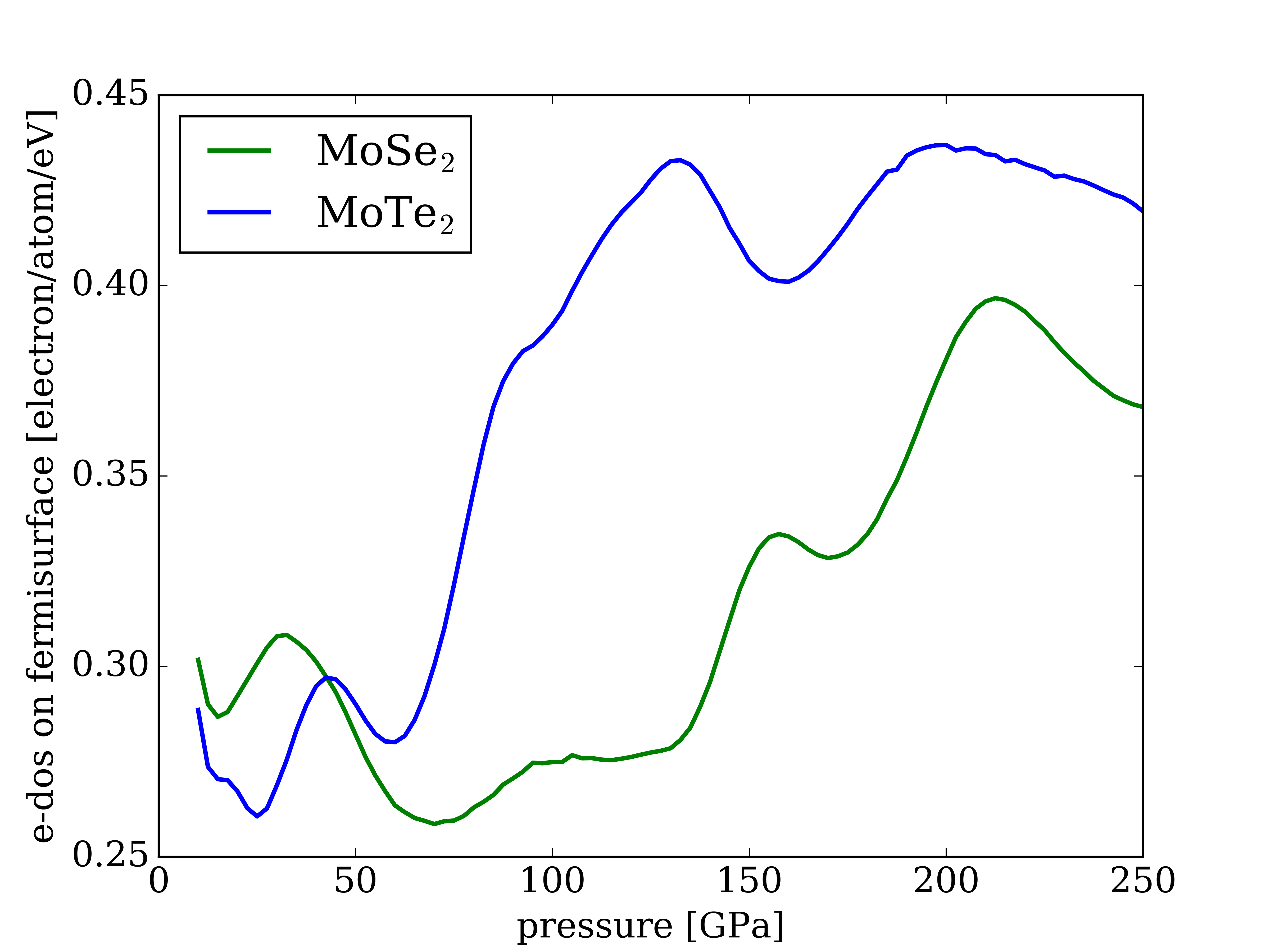}
    \caption{Pressure evolution of the electronic density of states at the Fermi level of structure \textit{I4/mmm}. In order to make the curves smoother a Gaussian smearing with $\sigma$ = 0.06 eV was performed.}
    \label{fig:fdos_evolution}
  \end{figure}

\begin{figure}[ht]
    \centering
    \includegraphics[width=0.45\textwidth]{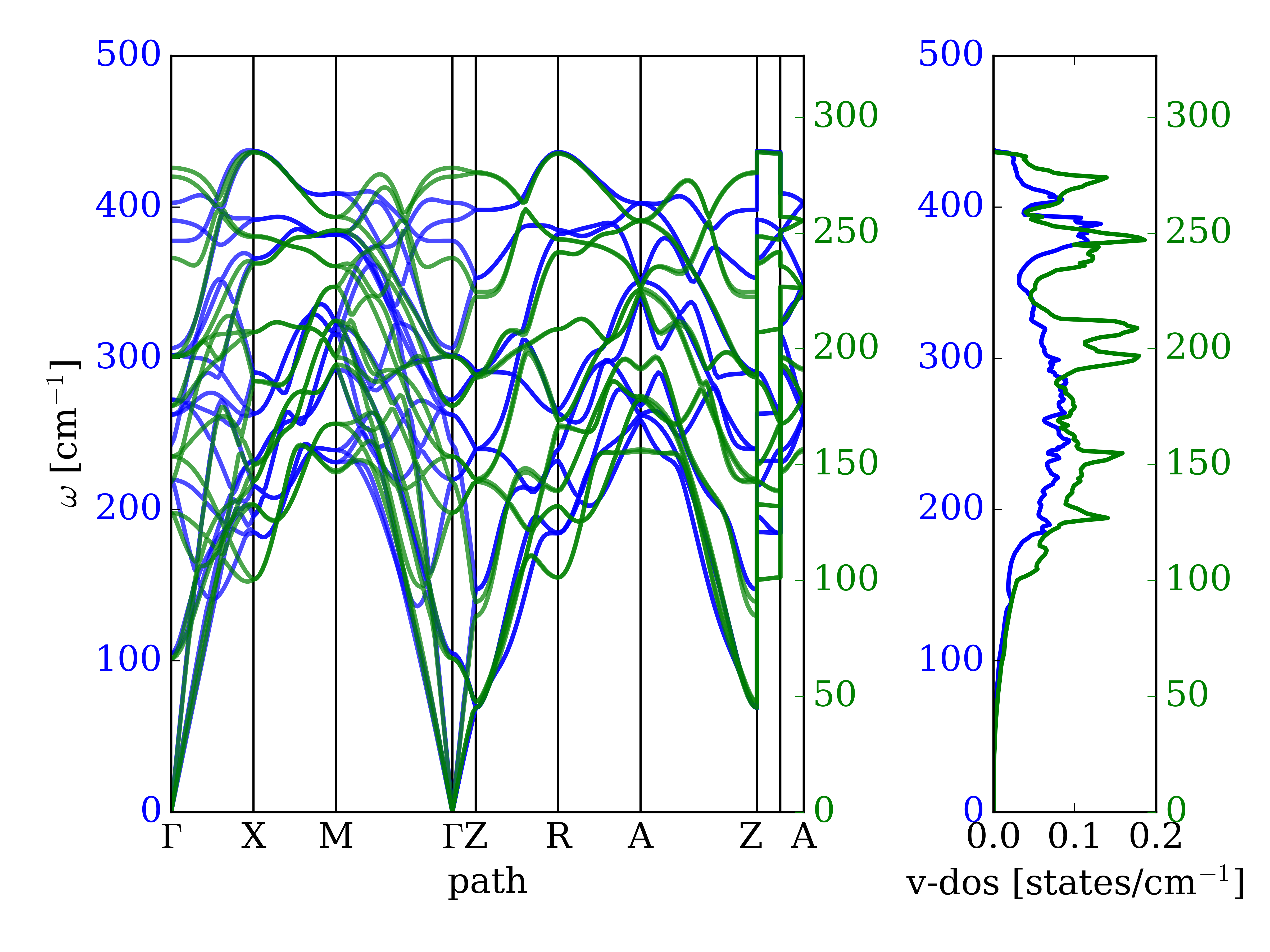}
    \caption{Phonon dispersion curves of the \textit{I4/mmm} structure. Blue and green lines are for MoSe$_2$ at 150 GPa and for MoTe$_2$ at 50 GPa, respectively. Last two sections are paths between X $\rightarrow$ R and M $\rightarrow$ A.}
    \label{fig:phonon}
  \end{figure}

In Fig.\ref{fig:MoSe2_I4mmm_bandstructure} we show the electronic band structure and density of states of the \textit{I4/mmm} structure at 150 GPa. It is clear that the structure is metallic. The pressure evolution of the density of states at the Fermi level for this phase is shown in Fig.\ref{fig:fdos_evolution}. It can be seen that the metallicity becomes stronger above 150 GPa as more bands cross the Fermi level. The most important contribution to states at the Fermi level originates from Mo-d states, in particular from the $d_{xz}$ and $d_{yz}$ states.
Employing the Quantum Espresso code we calculated in the \textit{I4/mmm} phase at 150 GPa the electron-phonon coupling parameter and found a value of $\lambda \sim 0.42 $. Applying the Allen-Dynes formula\cite{allen1975} and assuming $\mu^*=0.1$ we predict a superconducting critical temperature of $T_c \approx 2$ K. Phonon dispersion curves are shown in Fig.\ref{fig:phonon}.
  
\subsection{MoTe$_2$}
\label{MoTe2}

\begin{figure}[ht]
    \centering
    \includegraphics[width=0.45\textwidth]{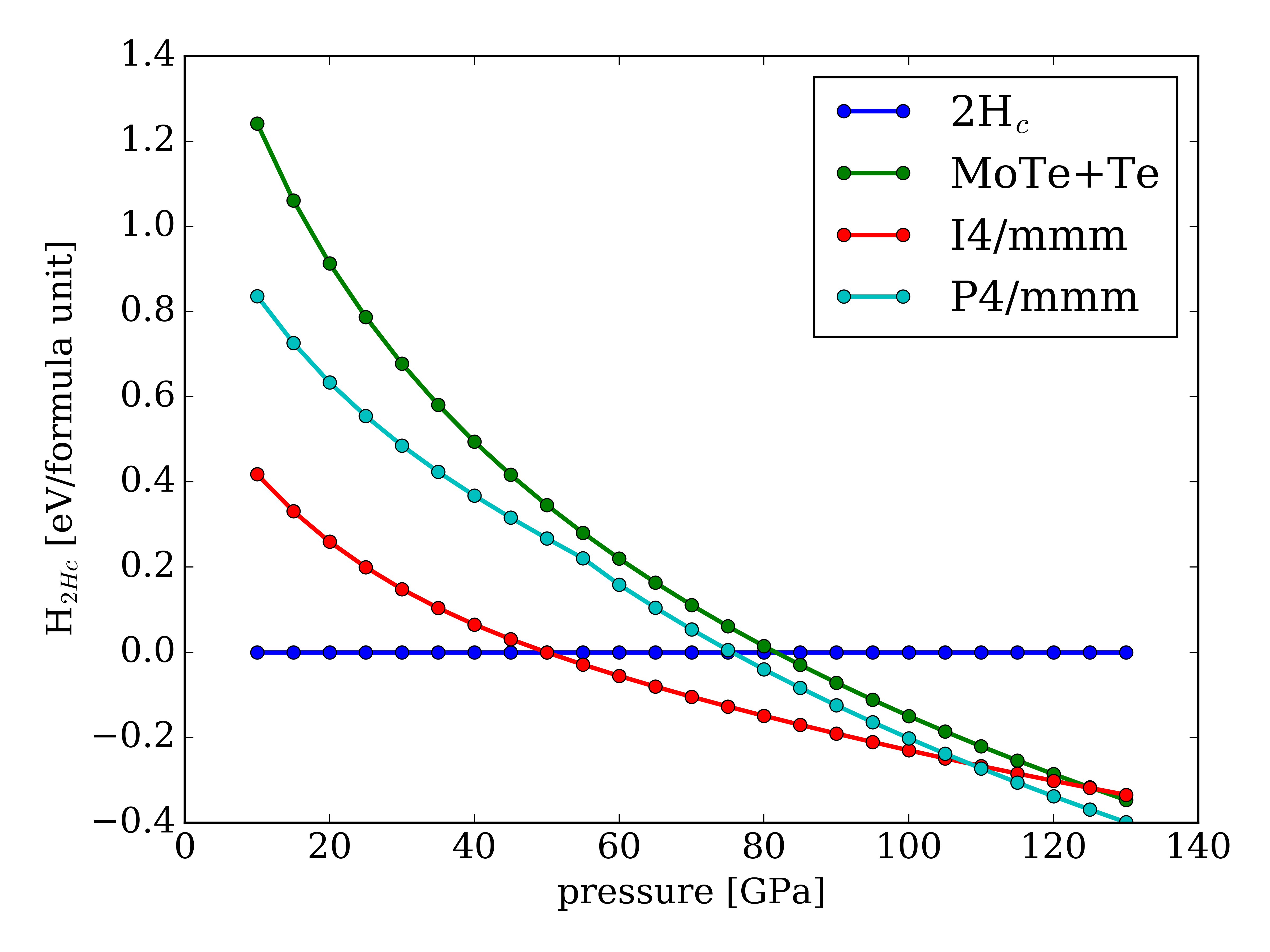}
    \caption{Enthalpy as function of pressure for MoTe$_2$ structures (calculated with VASP). Enthalpies are relative to the 2H$_c$ phase.}
    \label{fig:entcomp_decomp_MoTe2}
  \end{figure}

\begin{figure}[ht]
    \centering
    \includegraphics[width=0.45\textwidth]{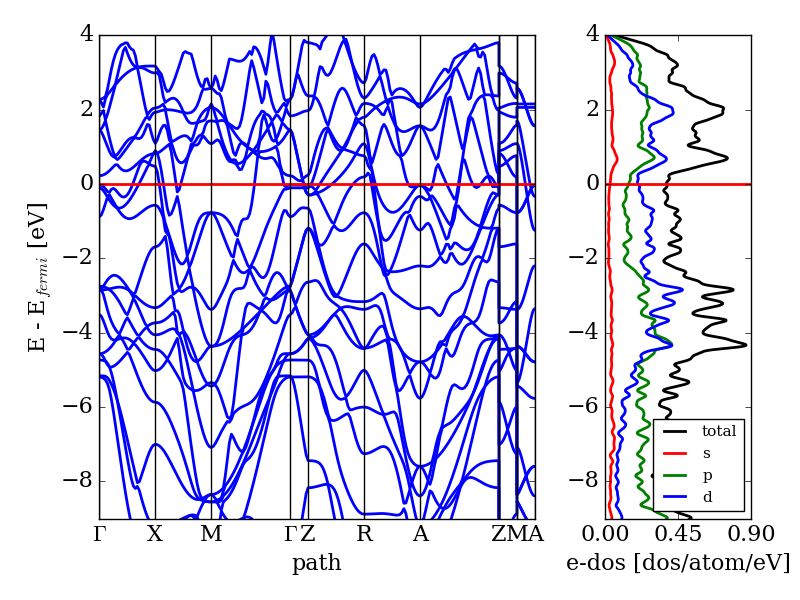}
    \caption{Band structure and projected density of states of the MoTe$_2$ - \textit{P4/mmm} phase at 110 GPa. Last two sections are paths between X $\rightarrow$ R and M $\rightarrow$ A.}
    \label{fig:MoTe2_P4mmm_bandstructure}
  \end{figure}

In MoTe$_2$ our structural search also found the same \textit{P4/mmm} and 
\textit{I4/mmm} structures as in the previous case. The enthalpies as function of pressure are shown in Fig.\ref{fig:entcomp_decomp_MoTe2}. In contrast to MoSe$_2$, however, here the transition to the \textit{I4/mmm} at 50 GPa preempts the transition to \textit{P4/mmm} as well as the instability with respect to chemical decomposition. We verified the possibility of the tentative $2H_c \rightarrow I4/mmm$ transition mechanism shown in Fig.\ref{fig_I4mmm} by simple simulation performing a gradual uniaxial compression of the 2H$_c$ phase of MoTe$_2$ at T=0. We found that such simulated compression indeed resulted in change of Mo atom coordination from six to eight and creation of the \textit{I4/mmm} phase. The transition is accompanied by a volume drop of 1.8 \%. Upon further compression the phase remains stable up to 110 GPa where the \textit{P4/mmm} structure takes over. The lattice parameters of the \textit{P4/mmm} phase at 110 GPa are $a = 3.041$ \AA~and $c = 9.471$ \AA, Mo atoms are on Wyckoff positions 2g (0.0, 0.0, 0.340) and Te atoms are on 1d (0.5, 0.5, 0.5), 2h (0.5, 0.5, 0.189) and 1a (0.0, 0.0, 0.0) positions. Phase transition from \textit{I4/mmm} to \textit{P4/mmm} is accompanied by volume drop of 1.2 \%. Electronic band structure and density of states are plotted in Fig.\ref{fig:MoTe2_P4mmm_bandstructure}, from which clear metallic character can be seen.

Since in case of MoTe$_2$ the \textit{P4/mmm} phase appears after the \textit{I4/mmm} where the eightfold coordination of Mo atoms already exists, it appears plausible that the kinetics of the transition $I4/mmm \rightarrow P4/mmm$ in MoTe$_2$ could be faster than that of direct transition 2H$_c \rightarrow P4/mmm$ in MoSe$_2$ and MoS$_2$ which requires an increase of coordination of Mo atoms. A speculative simple mechanism for the $I4/mmm \rightarrow P4/mmm$ transition, consisting of exchange of positions between Mo and chalcogen atoms is shown in Fig.\ref{fig_I4mmm}. In this respect we can speculate that the observed metastability of the 2H$_a$ phase up to 200 GPa in Ref.\cite{chi2015} could be possibly related to the high enthalpy of the \textit{I4/mmm} phase in MoS$_2$ which prevents it from acting as intermediate phase in creation of the \textit{P4/mmm} phase.

Assuming MoTe adopts the same CsCl structure as found in case of MoS$_2$ and MoSe$_2$ and Te adopts the bcc Te-V\cite{holzapfel1988} structure\footnote{This structure was found at pressure of 27 GPa in Ref.\cite{holzapfel1988} and we are not aware of experiments at higher pressure.}, MoTe$_2$ does not appear to be prone to chemical decomposition at any pressure between 0 and 130 GPa.

\begin{figure}[ht]
    \centering
    \includegraphics[width=0.45\textwidth]{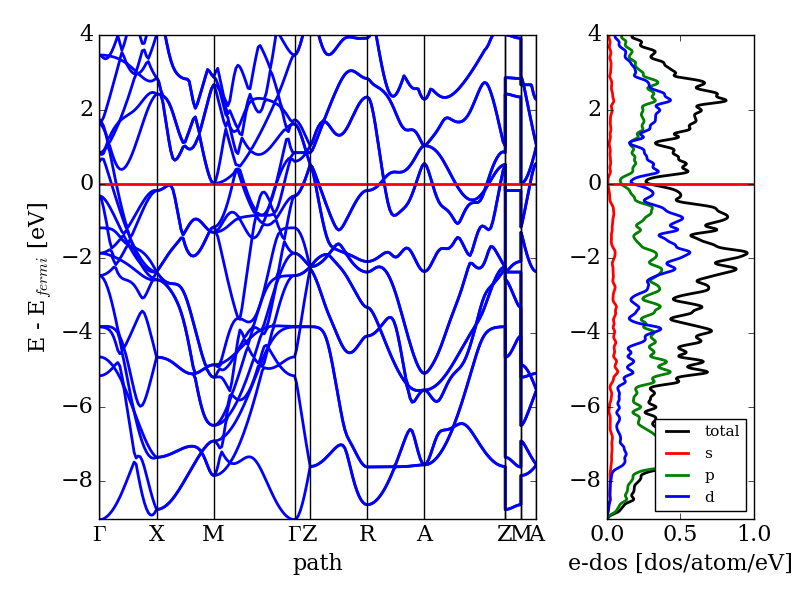}
    \caption{Band structure and projected density of states of the MoTe$_2$ - \textit{I4/mmm} phase at 50 GPa. Last two sections are paths between X $\rightarrow$ R and M $\rightarrow$ A.}
    \label{fig:MoTe2_I4mmm_bandstructure}
  \end{figure}

Lattice parameters of the \textit{I4/mmm} structure at 50 GPa (equilibrium condition with 2H$_c$ structure) are $a = 3.036$ \AA~and $c = 11.095$ \AA~with Mo atom on Wyckoff position 2a (0.0, 0.0, 0.0), and Te atom on 4e (0.5,  0.5, 0.145). The band structure and electronic density of states are shown in Fig. \ref{fig:MoTe2_I4mmm_bandstructure}. Projections of density of states show that the electronic states at the Fermi level are composed mostly of Mo-d orbitals (again mainly the $d_{xz}$ and $d_{yz}$ ones, see Fig.1 in Supplemental material\cite{Sup_mat}). While the \textit{I4/mmm} structure is metallic already at the transition pressure of 50 GPa, its metallicity substantially increases beyond 1 Mbar (see Fig. \ref{fig:fdos_evolution}). Using the same procedure as in case of MoSe$_2$ we predict at 50 GPa a coupling parameter $\lambda = 0.46$ resulting in superconducting critical temperature of $T_c \approx 2$ K. Phonon dispersion curves are shown in Fig.\ref{fig:phonon}.

Since the \textit{I4/mmm} structure becomes stable at relatively low pressure of 50 GPa, it is natural to ask whether it could be quenched to ambient pressure. In order to test this hypothesis we performed a short NPT MD run at $T=300$ K and $p=0$ GPa in a supercell consisting of 144 atoms. In this case we also added the vdW correction employing the Grimme DFT-D2 method\cite{grimme2006}. We observed after 4 ps a spontaneous transition to 2H structure with random stacking of layers, proceeding via mechanism essentially inverse to that described above for the $2H_c \rightarrow  I4/mmm$ transition (Fig.\ref{fig_I4mmm}). This shows that the \textit{I4/mmm} phase of MoTe$_2$ cannot be quenched to ambient pressure.
  
\section{Discussion and conclusions}
\label{conclusions}

Referring also to results found in Ref.\cite{kohulak2015} for MoS$_2$, we found three possible high-pressure structures to which the 2H MoX$_2$ compounds (2H$_c$ MoSe$_2$ or MoTe$_2$ and 2H$_a$ MoS$_2$) can transform, namely \textit{P4/mmm}, \textit{I4/mmm} and chemical decomposition. Comparing the three compounds, clear chemical trends can be seen for the post-2H phases. In case of MoS$_2$ chemical decomposition represents thermodynamically stable route since both \textit{P4/mmm} and \textit{I4/mmm} have higher and much higher enthalpy, respectively, and are only metastable. Decomposition could thus be possibly avoided only due to kinetic barriers. Moving towards heavier chalcogens from S to Se to Te the picture changes. For MoSe$_2$ and MoTe$_2$ chemical decomposition does not become thermodynamically stable in the whole pressure range in which we have studied them. Instead, \textit{P4/mmm} and in particular \textit{I4/mmm} become more favourable and eventually thermodynamically stable, resulting in transition 2H$_c \rightarrow P4/mmm$ in MoSe$_2$ at 118 GPa and 2H$_c \rightarrow I4/mmm$ in MoTe$_2$ at 50 GPa. 
Since the \textit{P4/mmm} phase becomes stable in MoTe$_2$ at 110 GPa, following $I4/mmm$, it appears likely that MoTe$_2$ might offer the easiest route to reach both post-2H phases in Mo-based TMD. We suggest that our predictions for new metallic high-pressure phases in MoSe$_2$ and MoTe$_2$ be tested experimentally. Besides pressure it might be necessary to work at elevated temperatures in order to overcome the presumably slow kinetics. The high-pressure layered structure \textit{I4/mmm} in both studied systems could possibly be interesting also as high-pressure lubricant, similarly to the case of MoS$_2$ at ambient pressure\cite{dallavalle2012}.

  \begin{acknowledgments}
    This work was supported by the Slovak Research and Development Agency under Contract No.~APVV-15-0496, by the VEGA project No.~1/0904/15 and by the project implementation 26220220004 within the Research \& Development Operational Programme funded by the ERDF. Part of the calculations were performed in the Computing Centre of the Slovak Academy of Sciences using the supercomputing infrastructure acquired in project ITMS 26230120002 and 26210120002 (Slovak infrastructure for high-performance computing) supported by the Research \& Development Operational Programme funded by the ERDF.
  \end{acknowledgments}


%

\newpage
\includepdf{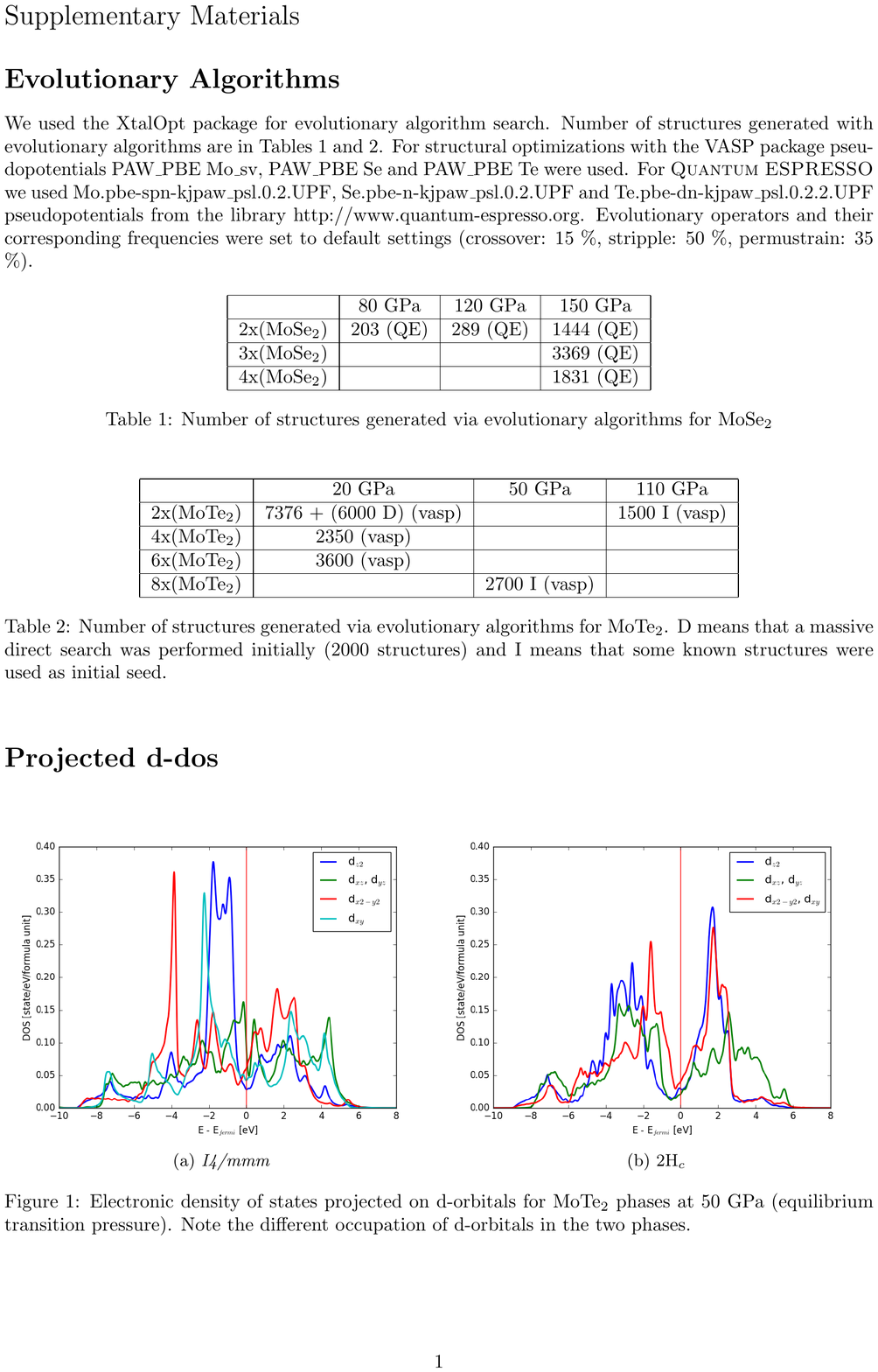}


\begin{thebibliography}{35}%
\makeatletter
\providecommand \@ifxundefined [1]{%
 \@ifx{#1\undefined}
}%
\providecommand \@ifnum [1]{%
 \ifnum #1\expandafter \@firstoftwo
 \else \expandafter \@secondoftwo
 \fi
}%
\providecommand \@ifx [1]{%
 \ifx #1\expandafter \@firstoftwo
 \else \expandafter \@secondoftwo
 \fi
}%
\providecommand \natexlab [1]{#1}%
\providecommand \enquote  [1]{``#1''}%
\providecommand \bibnamefont  [1]{#1}%
\providecommand \bibfnamefont [1]{#1}%
\providecommand \citenamefont [1]{#1}%
\providecommand \href@noop [0]{\@secondoftwo}%
\providecommand \href [0]{\begingroup \@sanitize@url \@href}%
\providecommand \@href[1]{\@@startlink{#1}\@@href}%
\providecommand \@@href[1]{\endgroup#1\@@endlink}%
\providecommand \@sanitize@url [0]{\catcode `\\12\catcode `\$12\catcode
  `\&12\catcode `\#12\catcode `\^12\catcode `\_12\catcode `\%12\relax}%
\providecommand \@@startlink[1]{}%
\providecommand \@@endlink[0]{}%
\providecommand \url  [0]{\begingroup\@sanitize@url \@url }%
\providecommand \@url [1]{\endgroup\@href {#1}{\urlprefix }}%
\providecommand \urlprefix  [0]{URL }%
\providecommand \Eprint [0]{\href }%
\providecommand \doibase [0]{http://dx.doi.org/}%
\providecommand \selectlanguage [0]{\@gobble}%
\providecommand \bibinfo  [0]{\@secondoftwo}%
\providecommand \bibfield  [0]{\@secondoftwo}%
\providecommand \translation [1]{[#1]}%
\providecommand \BibitemOpen [0]{}%
\providecommand \bibitemStop [0]{}%
\providecommand \bibitemNoStop [0]{.\EOS\space}%
\providecommand \EOS [0]{\spacefactor3000\relax}%
\providecommand \BibitemShut  [1]{\csname bibitem#1\endcsname}%
\let\auto@bib@innerbib\@empty
\bibitem [{\citenamefont {Wilson}\ and\ \citenamefont
  {Yoffe}(1969)}]{wilson1969}%
  \BibitemOpen
  \bibfield  {author} {\bibinfo {author} {\bibfnamefont {J.}~\bibnamefont
  {Wilson}}\ and\ \bibinfo {author} {\bibfnamefont {A.}~\bibnamefont {Yoffe}},\
  }\href {\doibase 10.1080/00018736900101307} {\bibfield  {journal} {\bibinfo
  {journal} {Advances in Physics}\ }\textbf {\bibinfo {volume} {18}},\ \bibinfo
  {pages} {193} (\bibinfo {year} {1969})},\ \Eprint
  {http://arxiv.org/abs/http://dx.doi.org/10.1080/00018736900101307}
  {http://dx.doi.org/10.1080/00018736900101307} \BibitemShut {NoStop}%
\bibitem [{\citenamefont {Gmelin}(1995)}]{gmelin1995}%
  \BibitemOpen
  \bibfield  {author} {\bibinfo {author} {\bibfnamefont {L.}~\bibnamefont
  {Gmelin}},\ }\href@noop {} {\emph {\bibinfo {title} {Gmelin Handbook of
  Inorganic and Organometallic Chemistry}}},\ Vol.\ \bibinfo {volume} {B 7-9}\
  (\bibinfo  {publisher} {Springer-Verlag, Berlin},\ \bibinfo {year} {1995})\
  p.~\bibinfo {pages} {16}\BibitemShut {NoStop}%
\bibitem [{\citenamefont {Mak}\ \emph {et~al.}(2010)\citenamefont {Mak},
  \citenamefont {Lee}, \citenamefont {Hone}, \citenamefont {Shan},\ and\
  \citenamefont {Heinz}}]{mak2010}%
  \BibitemOpen
  \bibfield  {author} {\bibinfo {author} {\bibfnamefont {K.~F.}\ \bibnamefont
  {Mak}}, \bibinfo {author} {\bibfnamefont {C.}~\bibnamefont {Lee}}, \bibinfo
  {author} {\bibfnamefont {J.}~\bibnamefont {Hone}}, \bibinfo {author}
  {\bibfnamefont {J.}~\bibnamefont {Shan}}, \ and\ \bibinfo {author}
  {\bibfnamefont {T.~F.}\ \bibnamefont {Heinz}},\ }\href {\doibase
  10.1103/PhysRevLett.105.136805} {\bibfield  {journal} {\bibinfo  {journal}
  {Phys. Rev. Lett.}\ }\textbf {\bibinfo {volume} {105}},\ \bibinfo {pages}
  {136805} (\bibinfo {year} {2010})}\BibitemShut {NoStop}%
\bibitem [{\citenamefont {Chang}\ and\ \citenamefont {Chen}(2011)}]{chang2011}%
  \BibitemOpen
  \bibfield  {author} {\bibinfo {author} {\bibfnamefont {K.}~\bibnamefont
  {Chang}}\ and\ \bibinfo {author} {\bibfnamefont {W.}~\bibnamefont {Chen}},\
  }\href {\doibase 10.1021/nn200659w} {\bibfield  {journal} {\bibinfo
  {journal} {ACS Nano}\ }\textbf {\bibinfo {volume} {5}},\ \bibinfo {pages}
  {4720} (\bibinfo {year} {2011})},\ \Eprint
  {http://arxiv.org/abs/http://dx.doi.org/10.1021/nn200659w}
  {http://dx.doi.org/10.1021/nn200659w} \BibitemShut {NoStop}%
\bibitem [{\citenamefont {Radisavljevic}\ \emph {et~al.}(2011)\citenamefont
  {Radisavljevic}, \citenamefont {Radenovic}, \citenamefont {Brivio},
  \citenamefont {Giacometti},\ and\ \citenamefont {Kis}}]{radisavljevic2011}%
  \BibitemOpen
  \bibfield  {author} {\bibinfo {author} {\bibfnamefont {B.}~\bibnamefont
  {Radisavljevic}}, \bibinfo {author} {\bibfnamefont {A.}~\bibnamefont
  {Radenovic}}, \bibinfo {author} {\bibfnamefont {J.}~\bibnamefont {Brivio}},
  \bibinfo {author} {\bibfnamefont {V.}~\bibnamefont {Giacometti}}, \ and\
  \bibinfo {author} {\bibfnamefont {A.}~\bibnamefont {Kis}},\ }\href {\doibase
  10.1038/nnano.2010.279} {\bibfield  {journal} {\bibinfo  {journal} {Nat
  Nano}\ }\textbf {\bibinfo {volume} {6}},\ \bibinfo {pages} {147} (\bibinfo
  {year} {2011})}\BibitemShut {NoStop}%
\bibitem [{\citenamefont {Wang}\ \emph {et~al.}(2012)\citenamefont {Wang},
  \citenamefont {Kalantar-Zadeh}, \citenamefont {Kis}, \citenamefont
  {Coleman},\ and\ \citenamefont {Strano}}]{wang2012}%
  \BibitemOpen
  \bibfield  {author} {\bibinfo {author} {\bibfnamefont {Q.~H.}\ \bibnamefont
  {Wang}}, \bibinfo {author} {\bibfnamefont {K.}~\bibnamefont
  {Kalantar-Zadeh}}, \bibinfo {author} {\bibfnamefont {A.}~\bibnamefont {Kis}},
  \bibinfo {author} {\bibfnamefont {J.~N.}\ \bibnamefont {Coleman}}, \ and\
  \bibinfo {author} {\bibfnamefont {M.~S.}\ \bibnamefont {Strano}},\ }\href
  {\doibase 10.1038/nnano.2012.193} {\bibfield  {journal} {\bibinfo  {journal}
  {Nat Nano}\ }\textbf {\bibinfo {volume} {7}},\ \bibinfo {pages} {699}
  (\bibinfo {year} {2012})}\BibitemShut {NoStop}%
\bibitem [{\citenamefont {Somoano}\ \emph {et~al.}(1973)\citenamefont
  {Somoano}, \citenamefont {Hadek},\ and\ \citenamefont
  {Rembaum}}]{somoano1973}%
  \BibitemOpen
  \bibfield  {author} {\bibinfo {author} {\bibfnamefont {R.~B.}\ \bibnamefont
  {Somoano}}, \bibinfo {author} {\bibfnamefont {V.}~\bibnamefont {Hadek}}, \
  and\ \bibinfo {author} {\bibfnamefont {A.}~\bibnamefont {Rembaum}},\ }\href
  {\doibase http://dx.doi.org/10.1063/1.1679256} {\bibfield  {journal}
  {\bibinfo  {journal} {The Journal of Chemical Physics}\ }\textbf {\bibinfo
  {volume} {58}},\ \bibinfo {pages} {697} (\bibinfo {year} {1973})}\BibitemShut
  {NoStop}%
\bibitem [{\citenamefont {Brown}(1966)}]{brown1966}%
  \BibitemOpen
  \bibfield  {author} {\bibinfo {author} {\bibfnamefont {B.~E.}\ \bibnamefont
  {Brown}},\ }\href {\doibase 10.1107/S0365110X66000513} {\bibfield  {journal}
  {\bibinfo  {journal} {Acta Crystallographica}\ }\textbf {\bibinfo {volume}
  {20}},\ \bibinfo {pages} {268} (\bibinfo {year} {1966})}\BibitemShut
  {NoStop}%
\bibitem [{\citenamefont {Clarke}\ \emph {et~al.}(1978)\citenamefont {Clarke},
  \citenamefont {Marseglia},\ and\ \citenamefont {Hughes}}]{clarke1978}%
  \BibitemOpen
  \bibfield  {author} {\bibinfo {author} {\bibfnamefont {R.}~\bibnamefont
  {Clarke}}, \bibinfo {author} {\bibfnamefont {E.}~\bibnamefont {Marseglia}}, \
  and\ \bibinfo {author} {\bibfnamefont {H.~P.}\ \bibnamefont {Hughes}},\
  }\href {\doibase 10.1080/13642817808245670} {\bibfield  {journal} {\bibinfo
  {journal} {Philosophical Magazine Part B}\ }\textbf {\bibinfo {volume}
  {38}},\ \bibinfo {pages} {121} (\bibinfo {year} {1978})},\ \Eprint
  {http://arxiv.org/abs/http://dx.doi.org/10.1080/13642817808245670}
  {http://dx.doi.org/10.1080/13642817808245670} \BibitemShut {NoStop}%
\bibitem [{\citenamefont {Aksoy}\ \emph {et~al.}(2006)\citenamefont {Aksoy},
  \citenamefont {Ma}, \citenamefont {Selvi}, \citenamefont {Chyu},
  \citenamefont {Ertas},\ and\ \citenamefont {White}}]{aksoy2006}%
  \BibitemOpen
  \bibfield  {author} {\bibinfo {author} {\bibfnamefont {R.}~\bibnamefont
  {Aksoy}}, \bibinfo {author} {\bibfnamefont {Y.}~\bibnamefont {Ma}}, \bibinfo
  {author} {\bibfnamefont {E.}~\bibnamefont {Selvi}}, \bibinfo {author}
  {\bibfnamefont {M.~C.}\ \bibnamefont {Chyu}}, \bibinfo {author}
  {\bibfnamefont {A.}~\bibnamefont {Ertas}}, \ and\ \bibinfo {author}
  {\bibfnamefont {A.}~\bibnamefont {White}},\ }\href {\doibase
  http://dx.doi.org/10.1016/j.jpcs.2006.05.058} {\bibfield  {journal} {\bibinfo
   {journal} {Journal of Physics and Chemistry of Solids}\ }\textbf {\bibinfo
  {volume} {67}},\ \bibinfo {pages} {1914 } (\bibinfo {year}
  {2006})}\BibitemShut {NoStop}%
\bibitem [{\citenamefont {Hromadov\'a}\ \emph {et~al.}(2013)\citenamefont
  {Hromadov\'a}, \citenamefont {Marto\ifmmode~\check{n}\else \v{n}\fi{}\'ak},\
  and\ \citenamefont {Tosatti}}]{hromadova2013}%
  \BibitemOpen
  \bibfield  {author} {\bibinfo {author} {\bibfnamefont {L.}~\bibnamefont
  {Hromadov\'a}}, \bibinfo {author} {\bibfnamefont {R.}~\bibnamefont
  {Marto\ifmmode~\check{n}\else \v{n}\fi{}\'ak}}, \ and\ \bibinfo {author}
  {\bibfnamefont {E.}~\bibnamefont {Tosatti}},\ }\href {\doibase
  10.1103/PhysRevB.87.144105} {\bibfield  {journal} {\bibinfo  {journal} {Phys.
  Rev. B}\ }\textbf {\bibinfo {volume} {87}},\ \bibinfo {pages} {144105}
  (\bibinfo {year} {2013})}\BibitemShut {NoStop}%
\bibitem [{\citenamefont {Bandaru}\ \emph {et~al.}(2014)\citenamefont
  {Bandaru}, \citenamefont {Kumar}, \citenamefont {Sneed}, \citenamefont
  {Tschauner}, \citenamefont {Baker}, \citenamefont {Antonio}, \citenamefont
  {Luo}, \citenamefont {Hartmann}, \citenamefont {Zhao},\ and\ \citenamefont
  {Venkat}}]{bandaru2014}%
  \BibitemOpen
  \bibfield  {author} {\bibinfo {author} {\bibfnamefont {N.}~\bibnamefont
  {Bandaru}}, \bibinfo {author} {\bibfnamefont {R.~S.}\ \bibnamefont {Kumar}},
  \bibinfo {author} {\bibfnamefont {D.}~\bibnamefont {Sneed}}, \bibinfo
  {author} {\bibfnamefont {O.}~\bibnamefont {Tschauner}}, \bibinfo {author}
  {\bibfnamefont {J.}~\bibnamefont {Baker}}, \bibinfo {author} {\bibfnamefont
  {D.}~\bibnamefont {Antonio}}, \bibinfo {author} {\bibfnamefont {S.-N.}\
  \bibnamefont {Luo}}, \bibinfo {author} {\bibfnamefont {T.}~\bibnamefont
  {Hartmann}}, \bibinfo {author} {\bibfnamefont {Y.}~\bibnamefont {Zhao}}, \
  and\ \bibinfo {author} {\bibfnamefont {R.}~\bibnamefont {Venkat}},\ }\href
  {\doibase 10.1021/jp410167k} {\bibfield  {journal} {\bibinfo  {journal} {The
  Journal of Physical Chemistry C}\ }\textbf {\bibinfo {volume} {118}},\
  \bibinfo {pages} {3230} (\bibinfo {year} {2014})},\ \Eprint
  {http://arxiv.org/abs/http://dx.doi.org/10.1021/jp410167k}
  {http://dx.doi.org/10.1021/jp410167k} \BibitemShut {NoStop}%
\bibitem [{\citenamefont {Chi}\ \emph {et~al.}(2014)\citenamefont {Chi},
  \citenamefont {Zhao}, \citenamefont {Zhang}, \citenamefont {Goncharov},
  \citenamefont {Lobanov}, \citenamefont {Kagayama}, \citenamefont {Sakata},\
  and\ \citenamefont {Chen}}]{chi2014}%
  \BibitemOpen
  \bibfield  {author} {\bibinfo {author} {\bibfnamefont {Z.-H.}\ \bibnamefont
  {Chi}}, \bibinfo {author} {\bibfnamefont {X.-M.}\ \bibnamefont {Zhao}},
  \bibinfo {author} {\bibfnamefont {H.}~\bibnamefont {Zhang}}, \bibinfo
  {author} {\bibfnamefont {A.~F.}\ \bibnamefont {Goncharov}}, \bibinfo {author}
  {\bibfnamefont {S.~S.}\ \bibnamefont {Lobanov}}, \bibinfo {author}
  {\bibfnamefont {T.}~\bibnamefont {Kagayama}}, \bibinfo {author}
  {\bibfnamefont {M.}~\bibnamefont {Sakata}}, \ and\ \bibinfo {author}
  {\bibfnamefont {X.-J.}\ \bibnamefont {Chen}},\ }\href {\doibase
  10.1103/PhysRevLett.113.036802} {\bibfield  {journal} {\bibinfo  {journal}
  {Phys. Rev. Lett.}\ }\textbf {\bibinfo {volume} {113}},\ \bibinfo {pages}
  {036802} (\bibinfo {year} {2014})}\BibitemShut {NoStop}%
\bibitem [{\citenamefont {{Chi}}\ \emph {et~al.}(2015)\citenamefont {{Chi}}
  \emph {et~al.}}]{chi2015}%
  \BibitemOpen
  \bibfield  {author} {\bibinfo {author} {\bibfnamefont {Z.}~\bibnamefont
  {{Chi}}} \emph {et~al.},\ }\href@noop {} {\bibfield  {journal} {\bibinfo
  {journal} {ArXiv e-prints}\ } (\bibinfo {year} {2015})},\ \Eprint
  {http://arxiv.org/abs/1503.05331} {arXiv:1503.05331 [cond-mat.supr-con]}
  \BibitemShut {NoStop}%
\bibitem [{\citenamefont {Kohul\'ak}\ \emph {et~al.}(2015)\citenamefont
  {Kohul\'ak}, \citenamefont {Marto\ifmmode~\check{n}\else \v{n}\fi{}\'ak},\
  and\ \citenamefont {Tosatti}}]{kohulak2015}%
  \BibitemOpen
  \bibfield  {author} {\bibinfo {author} {\bibfnamefont {O.}~\bibnamefont
  {Kohul\'ak}}, \bibinfo {author} {\bibfnamefont {R.}~\bibnamefont
  {Marto\ifmmode~\check{n}\else \v{n}\fi{}\'ak}}, \ and\ \bibinfo {author}
  {\bibfnamefont {E.}~\bibnamefont {Tosatti}},\ }\href {\doibase
  10.1103/PhysRevB.91.144113} {\bibfield  {journal} {\bibinfo  {journal} {Phys.
  Rev. B}\ }\textbf {\bibinfo {volume} {91}},\ \bibinfo {pages} {144113}
  (\bibinfo {year} {2015})}\BibitemShut {NoStop}%
\bibitem [{\citenamefont {Riflikov\'a}\ \emph {et~al.}(2014)\citenamefont
  {Riflikov\'a}, \citenamefont {Marto\ifmmode~\check{n}\else \v{n}\fi{}\'ak},\
  and\ \citenamefont {Tosatti}}]{riflikova2014}%
  \BibitemOpen
  \bibfield  {author} {\bibinfo {author} {\bibfnamefont {M.}~\bibnamefont
  {Riflikov\'a}}, \bibinfo {author} {\bibfnamefont {R.}~\bibnamefont
  {Marto\ifmmode~\check{n}\else \v{n}\fi{}\'ak}}, \ and\ \bibinfo {author}
  {\bibfnamefont {E.}~\bibnamefont {Tosatti}},\ }\href {\doibase
  10.1103/PhysRevB.90.035108} {\bibfield  {journal} {\bibinfo  {journal} {Phys.
  Rev. B}\ }\textbf {\bibinfo {volume} {90}},\ \bibinfo {pages} {035108}
  (\bibinfo {year} {2014})}\BibitemShut {NoStop}%
\bibitem [{\citenamefont {Zhao}\ \emph {et~al.}(2015)\citenamefont {Zhao},
  \citenamefont {Zhang}, \citenamefont {Yuan}, \citenamefont {Wang} \emph
  {et~al.}}]{zhao2015}%
  \BibitemOpen
  \bibfield  {author} {\bibinfo {author} {\bibfnamefont {Z.}~\bibnamefont
  {Zhao}}, \bibinfo {author} {\bibfnamefont {H.}~\bibnamefont {Zhang}},
  \bibinfo {author} {\bibfnamefont {H.}~\bibnamefont {Yuan}}, \bibinfo {author}
  {\bibfnamefont {S.}~\bibnamefont {Wang}},  \emph {et~al.},\ }\href {\doibase
  10.1038/ncomms8312} {\bibfield  {journal} {\bibinfo  {journal} {Nat Commun}\
  }\textbf {\bibinfo {volume} {6:7312}} (\bibinfo {year} {2015}),\
  10.1038/ncomms8312}\BibitemShut {NoStop}%
\bibitem [{\citenamefont {Soluyanov}\ \emph {et~al.}(2015)\citenamefont
  {Soluyanov}, \citenamefont {Gresch}, \citenamefont {Wang}, \citenamefont
  {Wu}, \citenamefont {Troyer}, \citenamefont {Dai},\ and\ \citenamefont
  {Bernevig}}]{soluyanov2015}%
  \BibitemOpen
  \bibfield  {author} {\bibinfo {author} {\bibfnamefont {A.~A.}\ \bibnamefont
  {Soluyanov}}, \bibinfo {author} {\bibfnamefont {D.}~\bibnamefont {Gresch}},
  \bibinfo {author} {\bibfnamefont {Z.}~\bibnamefont {Wang}}, \bibinfo {author}
  {\bibfnamefont {Q.}~\bibnamefont {Wu}}, \bibinfo {author} {\bibfnamefont
  {M.}~\bibnamefont {Troyer}}, \bibinfo {author} {\bibfnamefont
  {X.}~\bibnamefont {Dai}}, \ and\ \bibinfo {author} {\bibfnamefont {B.~A.}\
  \bibnamefont {Bernevig}},\ }\href {http://dx.doi.org/10.1038/nature15768}
  {\bibfield  {journal} {\bibinfo  {journal} {Nature}\ }\textbf {\bibinfo
  {volume} {527}},\ \bibinfo {pages} {495} (\bibinfo {year}
  {2015})}\BibitemShut {NoStop}%
\bibitem [{\citenamefont {Qi}\ \emph {et~al.}(2016)\citenamefont {Qi} \emph
  {et~al.}}]{Qi2016}%
  \BibitemOpen
  \bibfield  {author} {\bibinfo {author} {\bibfnamefont {Y.}~\bibnamefont {Qi}}
  \emph {et~al.},\ }\href {http://dx.doi.org/10.1038/ncomms11038} {\bibfield
  {journal} {\bibinfo  {journal} {Nat Commun}\ }\textbf {\bibinfo {volume}
  {7:11038}} (\bibinfo {year} {2016})}\BibitemShut {NoStop}%
\bibitem [{\citenamefont {Lonie}\ and\ \citenamefont
  {Zurek}(2011)}]{lonie2011}%
  \BibitemOpen
  \bibfield  {author} {\bibinfo {author} {\bibfnamefont {D.~C.}\ \bibnamefont
  {Lonie}}\ and\ \bibinfo {author} {\bibfnamefont {E.}~\bibnamefont {Zurek}},\
  }\href {\doibase http://dx.doi.org/10.1016/j.cpc.2010.07.048} {\bibfield
  {journal} {\bibinfo  {journal} {Computer Physics Communications}\ }\textbf
  {\bibinfo {volume} {182}},\ \bibinfo {pages} {372 } (\bibinfo {year}
  {2011})}\BibitemShut {NoStop}%
\bibitem [{\citenamefont {Giannozzi}\ \emph {et~al.}(2009)\citenamefont
  {Giannozzi} \emph {et~al.}}]{giannozzi2009}%
  \BibitemOpen
  \bibfield  {author} {\bibinfo {author} {\bibfnamefont {P.}~\bibnamefont
  {Giannozzi}} \emph {et~al.},\ }\href {http://www.quantum-espresso.org}
  {\bibfield  {journal} {\bibinfo  {journal} {Journal of Physics: Condensed
  Matter}\ }\textbf {\bibinfo {volume} {21}},\ \bibinfo {pages} {395502 (19pp)}
  (\bibinfo {year} {2009})}\BibitemShut {NoStop}%
\bibitem [{\citenamefont {Kresse}\ and\ \citenamefont
  {Hafner}(1993)}]{kresse1993}%
  \BibitemOpen
  \bibfield  {author} {\bibinfo {author} {\bibfnamefont {G.}~\bibnamefont
  {Kresse}}\ and\ \bibinfo {author} {\bibfnamefont {J.}~\bibnamefont
  {Hafner}},\ }\href {\doibase 10.1103/PhysRevB.47.558} {\bibfield  {journal}
  {\bibinfo  {journal} {Phys. Rev. B}\ }\textbf {\bibinfo {volume} {47}},\
  \bibinfo {pages} {558} (\bibinfo {year} {1993})}\BibitemShut {NoStop}%
\bibitem [{\citenamefont {Bl\"ochl}(1994)}]{blochl1994}%
  \BibitemOpen
  \bibfield  {author} {\bibinfo {author} {\bibfnamefont {P.~E.}\ \bibnamefont
  {Bl\"ochl}},\ }\href {\doibase 10.1103/PhysRevB.50.17953} {\bibfield
  {journal} {\bibinfo  {journal} {Phys. Rev. B}\ }\textbf {\bibinfo {volume}
  {50}},\ \bibinfo {pages} {17953} (\bibinfo {year} {1994})}\BibitemShut
  {NoStop}%
\bibitem [{\citenamefont {Kresse}\ and\ \citenamefont
  {Joubert}(1999)}]{kresse1999}%
  \BibitemOpen
  \bibfield  {author} {\bibinfo {author} {\bibfnamefont {G.}~\bibnamefont
  {Kresse}}\ and\ \bibinfo {author} {\bibfnamefont {D.}~\bibnamefont
  {Joubert}},\ }\href {\doibase 10.1103/PhysRevB.59.1758} {\bibfield  {journal}
  {\bibinfo  {journal} {Phys. Rev. B}\ }\textbf {\bibinfo {volume} {59}},\
  \bibinfo {pages} {1758} (\bibinfo {year} {1999})}\BibitemShut {NoStop}%
\bibitem [{\citenamefont {Perdew}\ \emph {et~al.}(1996)\citenamefont {Perdew},
  \citenamefont {Burke},\ and\ \citenamefont {Ernzerhof}}]{perdew1996}%
  \BibitemOpen
  \bibfield  {author} {\bibinfo {author} {\bibfnamefont {J.~P.}\ \bibnamefont
  {Perdew}}, \bibinfo {author} {\bibfnamefont {K.}~\bibnamefont {Burke}}, \
  and\ \bibinfo {author} {\bibfnamefont {M.}~\bibnamefont {Ernzerhof}},\ }\href
  {\doibase 10.1103/PhysRevLett.77.3865} {\bibfield  {journal} {\bibinfo
  {journal} {Phys. Rev. Lett.}\ }\textbf {\bibinfo {volume} {77}},\ \bibinfo
  {pages} {3865} (\bibinfo {year} {1996})}\BibitemShut {NoStop}%
\bibitem [{\citenamefont {Aksoy}\ \emph {et~al.}(2008)\citenamefont {Aksoy},
  \citenamefont {Selvi},\ and\ \citenamefont {Ma}}]{aksoy2008}%
  \BibitemOpen
  \bibfield  {author} {\bibinfo {author} {\bibfnamefont {R.}~\bibnamefont
  {Aksoy}}, \bibinfo {author} {\bibfnamefont {E.}~\bibnamefont {Selvi}}, \ and\
  \bibinfo {author} {\bibfnamefont {Y.}~\bibnamefont {Ma}},\ }\href {\doibase
  http://dx.doi.org/10.1016/j.jpcs.2008.03.020} {\bibfield  {journal} {\bibinfo
   {journal} {Journal of Physics and Chemistry of Solids}\ }\textbf {\bibinfo
  {volume} {69}},\ \bibinfo {pages} {2138 } (\bibinfo {year} {2008})},\
  \bibinfo {note} {study of Matter Under Extreme Conditions 2007Study of Matter
  Under Extreme Conditions 2007}\BibitemShut {NoStop}%
\bibitem [{\citenamefont {Monkhorst}\ and\ \citenamefont
  {Pack}(1976)}]{monkhorst1976}%
  \BibitemOpen
  \bibfield  {author} {\bibinfo {author} {\bibfnamefont {H.~J.}\ \bibnamefont
  {Monkhorst}}\ and\ \bibinfo {author} {\bibfnamefont {J.~D.}\ \bibnamefont
  {Pack}},\ }\href {\doibase 10.1103/PhysRevB.13.5188} {\bibfield  {journal}
  {\bibinfo  {journal} {Phys. Rev. B}\ }\textbf {\bibinfo {volume} {13}},\
  \bibinfo {pages} {5188} (\bibinfo {year} {1976})}\BibitemShut {NoStop}%
\bibitem [{\citenamefont {Bahn}\ and\ \citenamefont
  {Jacobsen}(2002)}]{bahn2002}%
  \BibitemOpen
  \bibfield  {author} {\bibinfo {author} {\bibfnamefont {S.~R.}\ \bibnamefont
  {Bahn}}\ and\ \bibinfo {author} {\bibfnamefont {K.~W.}\ \bibnamefont
  {Jacobsen}},\ }\href {\doibase 10.1109/5992.998641} {\bibfield  {journal}
  {\bibinfo  {journal} {Comput. Sci. Eng.}\ }\textbf {\bibinfo {volume} {4}},\
  \bibinfo {pages} {56} (\bibinfo {year} {2002})}\BibitemShut {NoStop}%
\bibitem [{Sup()}]{Sup_mat}%
  \BibitemOpen
  \href@noop {} {}\bibinfo {note} {See Supplemental Material (at the end of the article) 
  for more information about evolutionary searches and projected electronic
density of states}\BibitemShut {NoStop}%
\bibitem [{\citenamefont {Degtyareva}\ \emph {et~al.}(2005)\citenamefont
  {Degtyareva}, \citenamefont {Gregoryanz}, \citenamefont {Somayazulu},
  \citenamefont {Mao},\ and\ \citenamefont {Hemley}}]{degtyareva2005}%
  \BibitemOpen
  \bibfield  {author} {\bibinfo {author} {\bibfnamefont {O.}~\bibnamefont
  {Degtyareva}}, \bibinfo {author} {\bibfnamefont {E.}~\bibnamefont
  {Gregoryanz}}, \bibinfo {author} {\bibfnamefont {M.}~\bibnamefont
  {Somayazulu}}, \bibinfo {author} {\bibfnamefont {H.-k.}\ \bibnamefont {Mao}},
  \ and\ \bibinfo {author} {\bibfnamefont {R.~J.}\ \bibnamefont {Hemley}},\
  }\href {\doibase 10.1103/PhysRevB.71.214104} {\bibfield  {journal} {\bibinfo
  {journal} {Phys. Rev. B}\ }\textbf {\bibinfo {volume} {71}},\ \bibinfo
  {pages} {214104} (\bibinfo {year} {2005})}\BibitemShut {NoStop}%
\bibitem [{\citenamefont {Momma}\ and\ \citenamefont
  {Izumi}(2011)}]{momma2011}%
  \BibitemOpen
  \bibfield  {author} {\bibinfo {author} {\bibfnamefont {K.}~\bibnamefont
  {Momma}}\ and\ \bibinfo {author} {\bibfnamefont {F.}~\bibnamefont {Izumi}},\
  }\href {\doibase 10.1107/S0021889811038970} {\bibfield  {journal} {\bibinfo
  {journal} {Journal of Applied Crystallography}\ }\textbf {\bibinfo {volume}
  {44}},\ \bibinfo {pages} {1272} (\bibinfo {year} {2011})}\BibitemShut
  {NoStop}%
\bibitem [{\citenamefont {Allen}\ and\ \citenamefont
  {Dynes}(1975)}]{allen1975}%
  \BibitemOpen
  \bibfield  {author} {\bibinfo {author} {\bibfnamefont {P.~B.}\ \bibnamefont
  {Allen}}\ and\ \bibinfo {author} {\bibfnamefont {R.~C.}\ \bibnamefont
  {Dynes}},\ }\href {\doibase 10.1103/PhysRevB.12.905} {\bibfield  {journal}
  {\bibinfo  {journal} {Phys. Rev. B}\ }\textbf {\bibinfo {volume} {12}},\
  \bibinfo {pages} {905} (\bibinfo {year} {1975})}\BibitemShut {NoStop}%
\bibitem [{\citenamefont {Parthasarathy}\ and\ \citenamefont
  {Holzapfel}(1988)}]{holzapfel1988}%
  \BibitemOpen
  \bibfield  {author} {\bibinfo {author} {\bibfnamefont {G.}~\bibnamefont
  {Parthasarathy}}\ and\ \bibinfo {author} {\bibfnamefont {W.~B.}\ \bibnamefont
  {Holzapfel}},\ }\href {\doibase 10.1103/PhysRevB.37.8499} {\bibfield
  {journal} {\bibinfo  {journal} {Phys. Rev. B}\ }\textbf {\bibinfo {volume}
  {37}},\ \bibinfo {pages} {8499} (\bibinfo {year} {1988})}\BibitemShut
  {NoStop}%
\bibitem [{\citenamefont {Grimme}(2006)}]{grimme2006}%
  \BibitemOpen
  \bibfield  {author} {\bibinfo {author} {\bibfnamefont {S.}~\bibnamefont
  {Grimme}},\ }\href {\doibase 10.1002/jcc.20495} {\bibfield  {journal}
  {\bibinfo  {journal} {Journal of Computational Chemistry}\ }\textbf {\bibinfo
  {volume} {27}},\ \bibinfo {pages} {1787} (\bibinfo {year}
  {2006})}\BibitemShut {NoStop}%
\bibitem [{\citenamefont {Dallavalle}\ \emph {et~al.}(2012)\citenamefont
  {Dallavalle}, \citenamefont {S\"andig},\ and\ \citenamefont
  {Zerbetto}}]{dallavalle2012}%
  \BibitemOpen
  \bibfield  {author} {\bibinfo {author} {\bibfnamefont {M.}~\bibnamefont
  {Dallavalle}}, \bibinfo {author} {\bibfnamefont {N.}~\bibnamefont
  {S\"andig}}, \ and\ \bibinfo {author} {\bibfnamefont {F.}~\bibnamefont
  {Zerbetto}},\ }\href {\doibase 10.1021/la300871q} {\bibfield  {journal}
  {\bibinfo  {journal} {Langmuir}\ }\textbf {\bibinfo {volume} {28}},\ \bibinfo
  {pages} {7393} (\bibinfo {year} {2012})},\ \Eprint
  {http://arxiv.org/abs/http://dx.doi.org/10.1021/la300871q}
  {http://dx.doi.org/10.1021/la300871q} \BibitemShut {NoStop}%
\end{thebibliography}
\end{document}